\documentclass[prd,twocolumn,showkeys,floatfix,footinbib]{revtex4}
\usepackage{graphicx} 
\usepackage{amssymb} 
\usepackage{amsmath}
\usepackage{epsfig}

\def\bea{\begin{eqnarray}} 
\def\eea{\end{eqnarray}}

\newcommand{\arrowcom}[1]{\textbf{$\Longrightarrow$ #1}}
\newcommand{\trans}{\ensuremath{t}}

\begin{document}
\title{Unitarity Constraints on Semi-hard Jet Production in Impact Parameter Space} 
\author{T. C. Rogers}
\affiliation{Department of
Physics, Pennsylvania State University,\\ University Park, PA  16802,
USA}
\author{A. M. Sta\'sto}
\affiliation{Department of
Physics, Pennsylvania State University,\\ University Park, PA  16802,
USA}
\affiliation{H. Niewodnicza\'nski Institute of Nuclear Physics, Krak\'ow, Poland}
\author{M. I. Strikman}
\affiliation{Department of
Physics, Pennsylvania State University,\\ University Park, PA  16802,
USA }

\begin{abstract}
The perturbative QCD formula for minijet production consitutes an
important ingredient in models describing the total cross section and multiparticle production in hadron-hadron scattering at high energies. 
Using  arguments based on s-channel unitarity we set bounds on the minimum value of 
$p_\trans$ for which the leading twist minijet formula can be used.
For large impact parameters where correlations between partons appear to be small
we find that the minimum value of 
$p_\trans^c$ should be greater than $2.5 \,{\rm GeV}$ for LHC energies and greater than 
$3.5\, {\rm GeV}$ for cosmic ray energies of about $50 \,{\rm TeV}$. 
We also argue that for collisions with values of impact parameters typical for heavy particle production
the values of minimum $p_\trans$ are likely to be considerably larger.
We also analyze and quantify  the potential role of saturation effects in the gluon density.
We find that although saturation effects   alone are not sufficient to restore unitarity, they are likely to play an important role at LHC energies. 
\end{abstract}

\maketitle

\section{Introduction}
\label{intro}
  
The high energy hadronic interactions which will be studied at  the LHC, and the 
interaction of ultra-high energy cosmic rays in the atmosphere 
both involve QCD effects that are still poorly understood and 
largely unexplored experimentally.
Therefore, in extrapolating Monte Carlo simulations to these
very high energies, one often must resort to phenomenological 
models~\cite{Sjostrand:2006za,Bahr:2007ni,DPMjet,nexus,QGSjet,SIBYLL} based on fits to lower 
energy data.  Of particular interest  
is the contribution from jets with moderate relative 
transverse momentum $p_\trans$ (so-called mini-jets or semi-hard 
jets).  

The basic input to these models is the total inclusive minijet cross 
section in hadron-hadron scattering, typically calculated 
using the leading-twist perturbative QCD (pQCD) factorization 
formula. 
Schematically, the minijet cross section is the convolution
integral of a parton density for each colliding hadron and 
a  partonic cross section for two-to-two parton scattering,
\begin{multline}
\label{eq:minischem}
\sigma_{2jet}^{inc} \sim \int_{p_\trans^{c \, 2}}^{\infty} d\, p_\trans^{2} 
\frac{d \hat{\sigma}}{d p_\trans^2} \,
f_{h_1}(x_1,\mu^2) \otimes f_{h_2}(x_2,\mu^2), 
\end{multline}
where $\otimes$ denotes convolution integrals in longitudinal momentum 
fractions $x_1,x_2$ for each of the colliding hadrons.
The scale $\mu$ is the factorization scale 
which is usually set to be equal to $p_t$,
the transverse momentum.
Equation~(\ref{eq:minischem}) has to be 
regulated in the low momentum regime by introducing
a cutoff $p_\trans^c$.  Equation~(\ref{eq:minischem}) constitutes
the semi-hard part in most models of high energy hadron 
scattering.  The part which includes soft interactions 
must be modeled separately.

Since the integrand in Eq.~(\ref{eq:minischem}) is sharply 
peaked at very low transverse momenta, the minijet cross 
section is highly sensitive to the value of the cutoff 
$p_t^c$.
The  problem of
determining a minimum $p_\trans$ is generic to all the models which aim to  describe the particle production in hadron collisions
~\cite{Sjostrand:2006za,Bahr:2007ni,DPMjet,nexus,QGSjet,SIBYLL,DPM,Engel:2003ac}.

The true range of validity 
for $p_\trans^c$ remains a point of some controversy.
The leading twist pQCD expression for Eq.~(\ref{eq:minischem})
is most reliable when $p_\trans^c$ is large.  However, one
hopes to utilize the full power of pQCD for as wide
a range of kinematics as possible.

Imposing unitarity in impact parameter space 
provides a potential constraint on 
allowed values of $p_\trans^c$ at very 
high energies.  
Unlike in the case of the total cross section, however, the 
unitarity constraints cannot be imposed directly on
(\ref{eq:minischem}), which is an inclusive cross section proportional
to average jet multiplicity.  Therefore, less direct 
methods which take jet multiplicity into account are needed.
A parton model based picture of multiple interactions has
been shown to be a natural framework~\cite{Ametller:1987ru} with which to address 
the unitarity issue.  This approach 
has been applied, for example, in~\cite{Engel:2001mm} in the context 
of Monte Carlos for cosmic ray air showers.

In this paper, we propose a simple method 
for constraining the minimum value of $p_\trans^c$ 
at large impact parameters ($b \gtrsim 1.5$~fm)
for which the minijet formula in Eq.~(\ref{eq:minischem})
makes sense.    
Our method is based on s-channel unitarity of the amplitude in the impact parameter representation.
Our results are sensitive only to the transverse distribution of gluons 
in the nucleon which is measured in small-$x$ hard exclusive processes.
We produce 
 conservative bounds on $p_\trans^c$ --  
the true minimum $p_\trans^c$ should be  higher
than what we find.  
In this paper we restrict attention to  large impact parameters
where it is reasonable to neglect parton correlations.  
Apart from this, however, we avoid discussing  
any specific model of the underlying dynamics responsible 
for unitarization.  
Nevertheless, the 
approach we use in this paper is likely to be applicable at smaller 
impact parameters, given a particular model of parton 
correlations.

The lower bound we obtain on the 
transverse  momentum cutoff is at least $2.5 ~{\rm GeV}$ for  
center-of-mass pp collisions at $14$~TeV and impact parameters 
 $\gtrsim 1.5~ {\rm fm}$.
We also consider here much higher  energies, $\sqrt{s} \approx 50$~TeV
which are relevant to cosmic ray studies.
Here, we find that the
minimum cutoff probably becomes closer to $3.5 ~{\rm GeV}$ for the same values of 
impact parameters for $pp$ collisions.
We also discuss possibilities for extending the
analysis to smaller impact parameters.

The structure of the paper is as follows: in the next section 
we introduce the formula for minijet production. By 
investigating the integrand at fixed values of $p_\trans$ and 
the energy $\sqrt{s}$ we constrain the range in the 
rapidities and $x$ values that  dominates the integral.  
On the basis of the saturation scale from the Golec-Biernat 
and Wusthoff model (GBW) we estimate the role of the 
saturation corrections in the gluon density at the 
Tevatron and at the LHC.

In Sec. III we develop a general
method based on unitarity arguments for constraining 
the minimum value of the $p_\trans$ for the minijet formula to be 
used.  

In Sec. IV we give a general discussion of the radius 
of the interaction in the context of the eikonal model.
We make comparisons here to the interaction radius for 
deep inelastic scattering (DIS)

Finally, in the appendix we include the formulae for the dependence of the 
mass parameter on $x$ and the scale  in the profile function.

\section{Minijet cross section}
We begin by writing Eq.~(\ref{eq:minischem}) more explicitly.
The inclusive minijet cross section in hadron-hadron scattering
is typically calculated using the leading-twist perturbative QCD (pQCD) 
factorization formula,
\begin{equation}
\label{minijets}
\begin{split}
\sigma_{2jet}^{inc}(s,p_{t}^{\rm c}) = \sum_{i,j,k,l} \frac{K}{1 + \delta_{kl}} 
\int d \, x_1 d \, x_2 \int d \, p_{t}^{2} \times \\ \times \,
\frac{d \hat{\sigma}_{i j \rightarrow k l}}{d p_{t}^{2}} \,
f_i (x_1 , \mu^{2})\, f_j (x_2 , \mu^{2})\, \theta(p_\trans - p_{t}^c)\; .
\end{split}
\end{equation}

Here, the differential cross section 
$d\hat{\sigma}/dp_\trans^2$ is for two-to-two parton
scattering.  In this work we calculate it at the lowest order in pQCD.  
The standard leading twist parton distribution function, $f_j(x,\mu^{2})$
is evaluated at a hard scale $\mu^{2}$ which is typically chosen to be
the transverse momentum squared, $p_\trans^{2}$.
The factor $1/(1 + \delta_{kl})$ takes into account necessary symmetry factors
when the outgoing partons are identical.
The $K$ is a factor used to correct for higher order terms, and is 
fixed to $2$ in many models. 
In NLO treatments of single jets, $K$ is found to be closer 
to $1.0$.  
However since processes where at least three jets are produced appear 
to be dominant  at these energies $K=1$ corresponds to
 effective $K \sim 1.5$ for our observable.
In keeping with our goal of maintaining conservative 
estimates of unitarity bounds, we take this value for the $K$ factor.\\

The kinematic limit for the $x_1,x_2$ integrations is
\begin{equation}
x_1 x_2 s >  4 (p_\trans^c)^2 \; .
\label{eq:ineq}
\end{equation}
As we will see shortly, in the UHE limit and at small 
$p_\trans$, the dominant range of the  integration in 
(\ref{minijets}) is over small values of $x$ where the 
gluon density, $g(x,p_\trans^2)$ dominates.  A continuing 
source of uncertainty is the lowest value of the  $p_{T}^c$ for 
which Eq.~(\ref{minijets}) makes sense.
Since the cutoff is 
in the regime where the integrand  
begins to spike, small changes in the cutoff can 
lead to  dramatic changes in the integrated cross section, 
Eq.~(\ref{minijets}).
This is illustrated in Fig.~\ref{minijet} where the integrated minijet 
cross section is shown as a function of the c.m.s. energy for different values 
of the $p_\trans^{\rm c}$. We see a drastic change in the 
normalization, although the shape and the rise of the cross 
section with the energy is universal.
This sensitivity to $p_\trans^c$ is a symptom of our limited understanding of 
physics in transitional region between small values of $p_\trans$, where the 
distribution in transverse momentum must be modeled, and large values where
Eq.~(\ref{minijets}) is a good approximation.

\subsection{The Kinematic Range of Validity for the 
Minijet Formula}

\begin{figure}
\epsfig{file=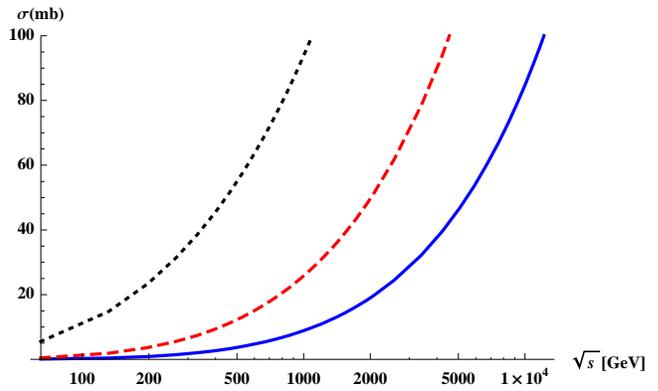,scale=.8}
\caption{Inclusive minijet cross section evaluated from (\ref{minijets}) 
for three different cutoff values $p_\trans^c=1.5$ (dotted line),  $p_\trans^c=2.5$ 
(dashed line) and $p_\trans^c=3.5 {\rm GeV}$ (solid line).  The $K$ factor is set to one in the plots.
Parton distributions are CTEQ6M.}
\label{minijet}
\end{figure}

In order to gain an intuition about the structure of 
the minijet cross-section at very high energies, it is useful
to study the integrand of Eq.~(\ref{minijets}) as a function of 
the rapidities $y_1,y_2$ of the jets.  
We show a contour plot in Fig.~\ref{minijet} for the integrand 
versus $y_1$ and $y_2$, related to $x_1$ and $x_2$ via the equation,
 $$
 x_{1,2} = \frac{p_\trans}{\sqrt{s}} (e^{\pm y_1}+e^{\pm y_2}) \; .
 $$
We use CTEQ6M \cite{Pumplin:2002vw} parton distribution sets for evaluation of  Eq.~(\ref{minijets}).
The smallest contours represent the maxima of the integrand.
To get an idea of the important regions of the integrand, we have drawn 
solid thick-red curves along the contours marking half of the peak value.
Plots (a) and (b) show the result of calculations done at $1.96$~TeV and 
$14$~TeV respectively.  In both cases, $p_t$ is fixed 
at $2.5$~GeV.  Note that as $\sqrt{s}$ increases, the peaks
begin to form long ridges that extend toward the lower-left and 
upper-right corners of the plots where $y_1$ and $y_2$ have the same sign.
The ridges also expand if we decrease $p_t$, as shown in plot (c)
where $p_t = 1.5$~GeV, still with $\sqrt{s} = 14$~TeV.  

For figure~\ref{minijets}(a) the peaks are located at rapidities,
$$
  y_1 \simeq - y_2 \simeq \pm 1.2 \, ,
$$
and so the $x$ values probed in the collision with this configuration are approximately equal
  $x_1\simeq x_2 \simeq 0.005$.   Still, at rapidities
 $$
  |y_1|\simeq4, |y_2|\simeq2, \; \; {\rm and} \; \;  {\rm sign}\,y_1={\rm sign}\,y_2 \; ,
 $$
the integrand is rather large. Such a configuration corresponds to asymmetric 
values of $x_1,x_2$,
$$
 x_{1,2} \simeq 2 \times 10^{-4} \ll x_{2,1} \simeq 0.08 \; .
$$
At the maximum LHC energy -- $14$~TeV --  plotted in figure (b),
the central values are,
\arrowcom{$x_1 \leftrightarrow x_2$}
$$
x_{1} \simeq x_{2} \simeq 0.0008,
$$
but the ridges extend toward values of rapidity,
$$
  |y_1|\simeq6, |y_2|\simeq4, 
$$
corresponding to 
 $
 x_{1,2} \simeq 4 \times 10^{-6} \ll x_{2,1} \simeq 0.08 \; .
 $
Therefore, both aligned and anti-aligned jets become important.

In the regions where $y_1$ and $y_2$ are both very large and 
of the same sign, one of the values of $x$ becomes very small.
The is the regime where one of the gluon densities begins to blow
up, and one expects non-linear effects (saturation) to come into play, see for example \cite{Gribov:1984tu},\cite{Iancu:2003xm}.
For a given value of $x$ 
saturation models predict that 
$g(x,\mu)$ becomes  saturated
for values of $\mu$ below the saturation scale, $Q_s$. 
Studies of deep inelastic scattering using 
various versions of the pQCD dipole 
picture~\cite{GolecBiernat:1998js,GolecBiernat:1999qd,Munier:2001nr,Rogers:2003vi,Kowalski:2003hm} have resulted in 
rough bounds on where saturation effects should begin to become significant.
To illustrate the sensitivity of the integrand to the saturation effects  in the gluon density
 we superimpose the saturation scale from the 
GBW model \cite{GolecBiernat:1998js},\cite{GolecBiernat:1999qd}  onto our contour plots.
The blue lines in the corners of the contour figures correspond to the saturation line\footnote{Note that, usually saturation line is referred to as line in $(x,\mu)$ space where $\mu$ is the typical scale.
Here, even though the scale $p_t$ is fixed in the contour plots we still have the line, since $x_{1,2}$ are functions of rapidities $y_1,y_2$.}
 from the GBW model.
These lines   mark off the  regions of saturation in 
the upper right-hand and lower left-hand corners of the figures.
Values of $y_1,y_2$ that lie within the blocks marked off by the saturation line indicate
regions where either $g(x_1,p_t)$ or $g(x_2,p_t)$ is likely to 
be very sensitive to saturation effects since either $Q_s(x_1)$ or $Q_s(x_2)$ are larger than the fixed value of $p_t$.  
The percentage of the integral affected by the saturation corrections is about $\sim 11 \%$ for the $p_t=2.5 \, {\rm GeV}$ and $\sim 53 \%$ for $p_t=1.5 \, {\rm GeV}$ for the LHC energy.
We did not show the saturation lines for the Tevatron energy
since for this choice of $p_t=2.5 \; {\rm GeV}$ the saturation effects are very small. 
We note, however, that 
the model used in~\cite{Rogers:2003vi} allows $x_1,x_2$ to go to somewhat smaller values 
without actually violating unitarity.

\begin{figure*}
\centering
  \begin{tabular}{c@{\hspace*{5mm}}c}
    \includegraphics[scale=0.3]{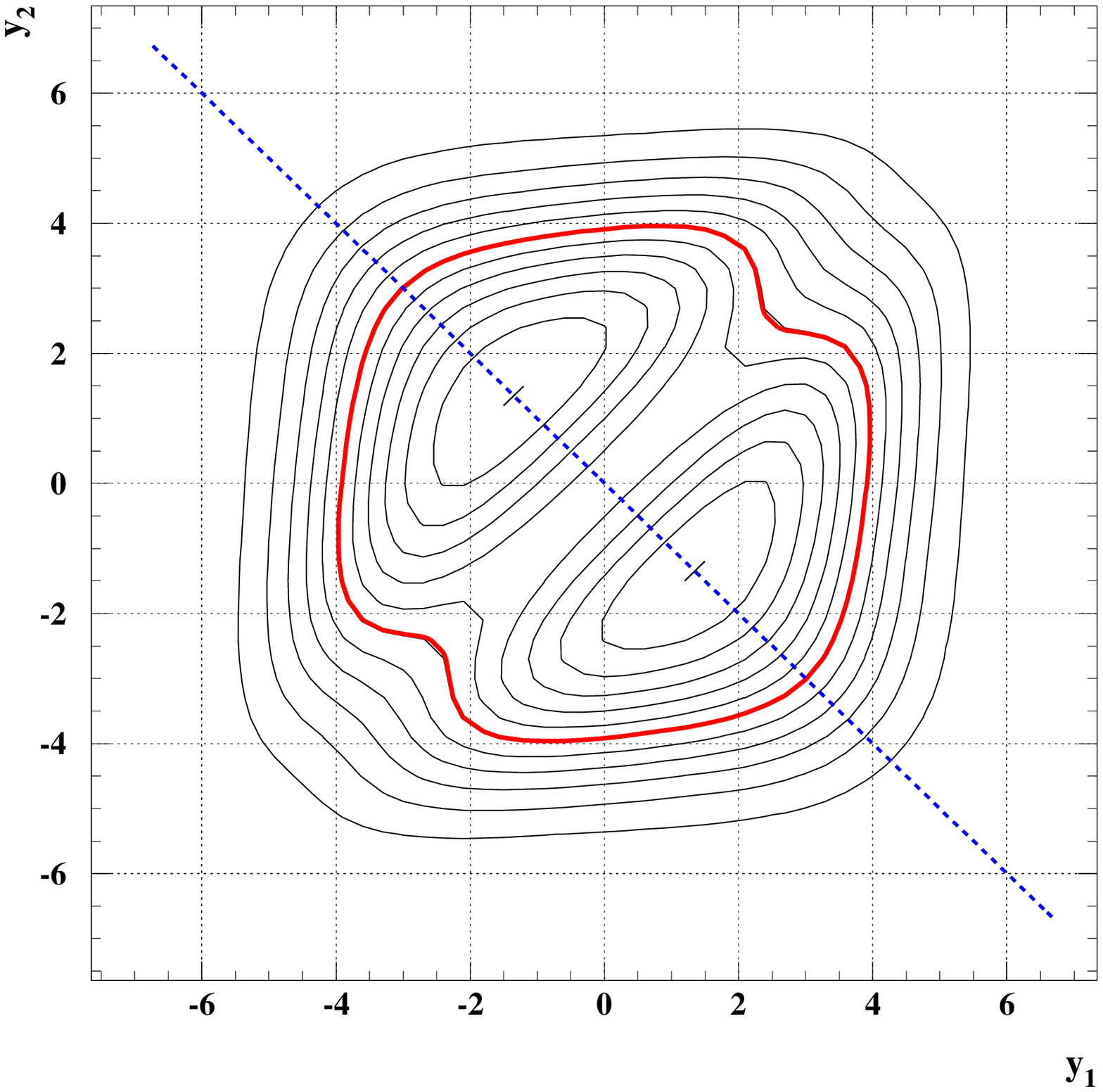}
    &
    \includegraphics[scale=0.3]{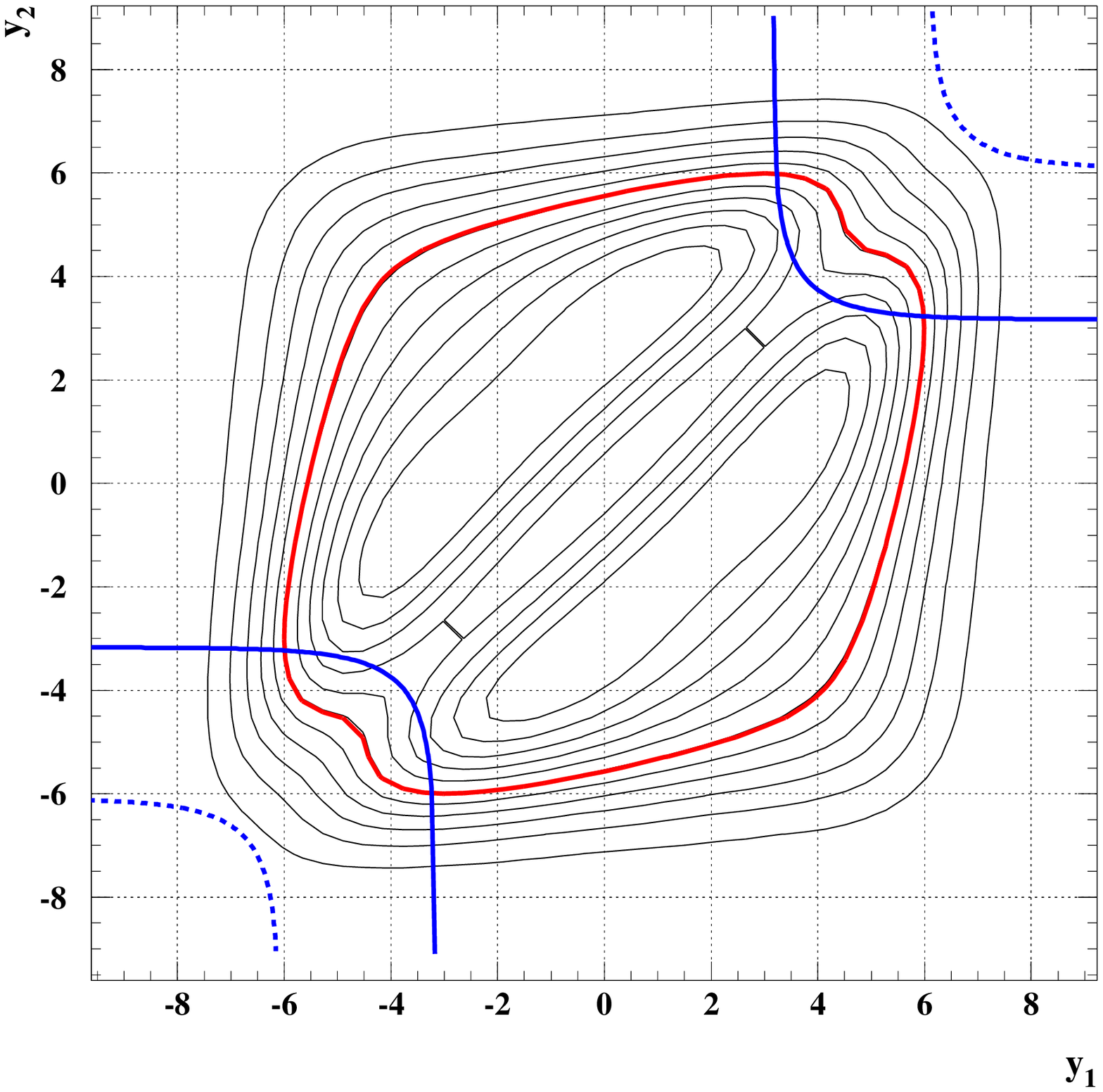}
  \\
  (a) & (b)
  \\[3mm]
  \multicolumn{2}{c}{
    \includegraphics[scale=0.3]{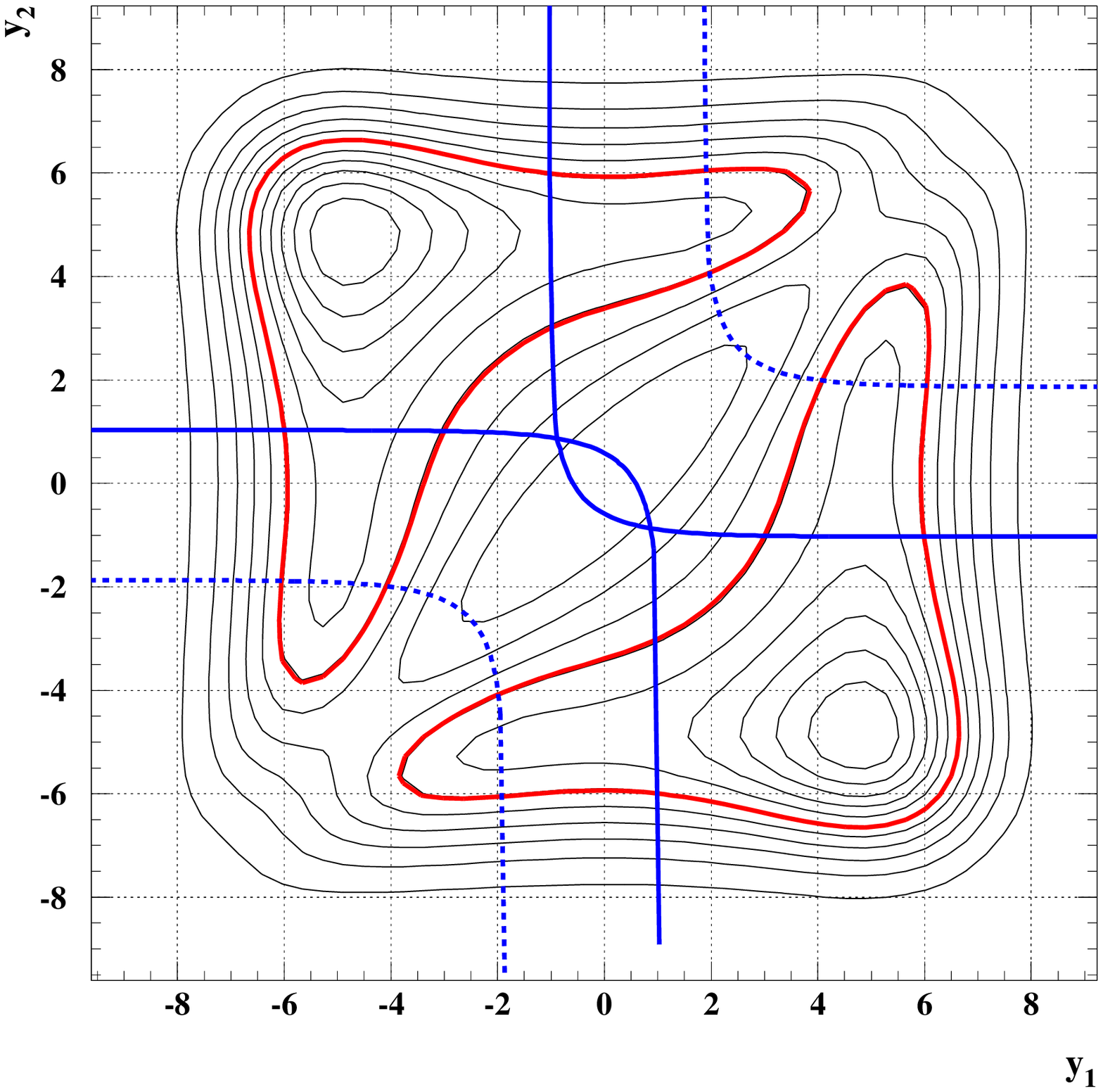}
  }
  \\
  \multicolumn{2}{c}{(c)}
  \end{tabular}
\caption{Rapidity distribution of the integrand in the minijet 
cross section formula for (a)$\sqrt{s}=1.96$~TeV, $p_\trans=2.5$~GeV, (b)
$\sqrt{s}$=14 TeV, $p_\trans=2.5$~GeV and (c) $\sqrt{s}$=14 TeV, $p_\trans=1.5$~GeV.
The calculation is done for the gluon-gluon  channel only.
The thick - red contour is where the integrand 
is $50\%$ of the maximum value.  The corner regions marked 
off by the dotted blue lines indicate where saturation effects may 
be present according to the GBW model.  The solid blue line is 
the result of rescaling the saturation scale by a 
factor of 9/4 as discussed in the text.}
\label{range_plots}
\end{figure*}

When evaluating the saturation effects one has to take into account the difference between the saturation scale 
for the quark and gluon dipoles. Since here the dominant process is $gg\rightarrow gg$ scattering the saturation
scale  is
enhanced by an additional color factor $C_A/C_F = 9/4$ relative~\cite{colorfactor} to 
the  case for the quark scattering.  We therefore include the corresponding rescaled saturation region
in Fig.~\ref{range_plots}, marked off by the solid blue line (as opposed to the dashed-blue which corresponds to the original unrescaled value of the saturation scale from the GBW model).  

Furthermore, our analysis has only
considered the leading $gg\rightarrow gg$ contribution so far. 
In the calculation of the total inclusive minijet cross section 
we include contributions from all channels. The $gg\rightarrow gg$ 
channel is of course dominant, it is about $60-70\%$ of the total 
value depending on the choice of the  parton density and the c.m.s. 
energy. The $qg\rightarrow qg$ contribution however, is non-negligible
 and constitutes about $30\%$ of the minijet cross section. It is 
particularly important for asymmetric configurations of the minijets. 
In this case the large contribution
comes from the region where the quark density is evaluated at 
large $x$ and the gluon density is evaluated at very small values of  $x$. 


Figures~\ref{range_plots}(a-b) demonstrate that gluon saturation effects will
likely be significant at LHC energies and higher.
However, we stress that other effects (e.g. multiple  scattering)
are  equally important for unitarizing the $pp$ cross section. 
In fact the multiple hadron scattering effects are  likely to become 
dominant at the lower energies, in a regime where 
the saturation of the gluon density is not very large.
In this paper, we take a very general approach that does not
distinguish the underlying physics responsible for unitarization.


\section{Impact Parameter Dependence of the Minijet Cross Section}
\label{sec:eikonal}

In this section we review the  basic 
elements needed to discuss unitarity in impact 
parameter space.

\subsection{Basic Formulation}

Following standard steps the 
total, elastic and inelastic cross sections can 
be written at high energies as,
\begin{eqnarray}
\label{eq:pp_profile}
\sigma_{tot}(s) & = & 2 \int d^2 {\bf b} \, {\rm Re} \, \Gamma(s,b), \label{eq:sigtot} \\
\sigma_{el}(s) & = & \int d^2 {\bf b} \, \left| \Gamma(s,b) \right|^2, \label{eq:sigel} \\
\sigma_{inel}(s) & = &\int d^2 {\bf b} \left( 2\,{\rm Re}  \,\Gamma(s,b) 
-  \left| \Gamma(s,b) \right|^2 \right) \; ,\label{eq:siginel} \,
\end{eqnarray}
where the profile function $\Gamma(s,b)$ is the Fourier transform in 
impact parameter of the  scattering amplitude for two-to-two process
\begin{equation}
\Gamma(s,b) = \frac{1}{2 i s(2 \pi)^2} \int d^2{\bf q} \,e^{i {\bf q} \cdot {\bf b}} A(s,t). 
\end{equation}
Assuming that the amplitude is dominantly imaginary 
(which is appropriate at high energies), the unitarity 
constraint on $\Gamma(b,s)$ is,
$$
\Gamma(s,b) \leq 1\;.
$$

\subsection{Impact Parameter Dependence for Hard Scattering}
\label{DIS}

We appeal directly to experimental data to obtain 
the impact parameter dependence of the hard collisions.
We use the generalized gluon distribution function extracted 
directly from $J/\Psi$ electroproduction production~\cite{Frankfurt:2002ka}.

The generalized gluon distribution is related 
to the standard gluon distribution function and a gluon 
form factor ${ F}_g(x,t,\mu^2)$ through the following defining relation
\begin{equation}
xf_g(x,t,\mu^2) = xf_g(x,\mu^2) \, { F}_g(x,t,\mu) \; ,
\label{eq:gpdf}
\end{equation}
where
\begin{equation}
{ F}_g(x,t=0,\mu) =1 \; .
\label{eq:normfg}
\end{equation}
The Fourier transform of the gluonic form factor gives the profile in impact parameter space
\begin{equation}
\label{eq:fftransform}
 {\cal F}_g(x,\rho,\mu)  =  \int d^2 {\bf \Delta} \, { F}_g(x,t,\mu) \, 
e^{-i {\bf \Delta} \cdot {\bf \rho}}, \hspace*{0.5cm} t=-\Delta^2 \; ,
\end{equation}
where the integration over $\rho$  is 2-dimensional. 
We have the normalization condition
$$
\int d^2 {\bf \rho}\,{\cal F}_g(x,\rho,\mu)= 1 \; ,
$$
which is a trivial consequence of the previous condition (\ref{eq:normfg}).
Interpreting ${\cal F}_g(x,\rho,\mu)$ as the 
transverse spatial spread of hard partons, we may write Eq.~(\ref{minijets}) 
in the form,
$$
\label{eq:N_norm}
\sigma_{2 jet}^{inc}(s,p_\trans^c) = \int d^2 {\bf b} \,\,  {\cal N}_2(b,s,p_\trans^c),
$$
where we have defined,
\begin{multline}
{\cal N}_2(b,s,p_\trans^c) = \sum_{k,l} \frac{K}{1 + \delta_{k,l}} \times 
\int_{0}^{1} d \, x_1 d \, x_2 \int d \, p_{T}^{2} \times \\ \times 
\frac{d \hat{\sigma}_{i j \rightarrow k l}}{d p_{\trans}^{2}} 
f_g (x_1 , p_\trans^{2}) \times \\ f_g (x_2 , p_\trans^{2}) P_2(b,x_1,x_2,p_\trans) \theta(p_\trans - p_\trans^c),
\label{eq:nsemcoll2}
\end{multline}
with an over-lap function given by,
\begin{multline}
P_2(b,x_1,x_2,\mu) = \\ \int d^2{\bf \rho_1} \,   \, {\cal F}_g(x_1,|{\bf \rho_1}|,\mu) \,  {\cal F}_g(x_2,|{\bf b}-{\bf \rho_1}|,\mu) \; .
\end{multline}
As usual, we use $\mu = p_t$ for the hard scale.
It follows from these definitions that ${\cal N}_2(b,s,p_\trans^c)$ should be
interpreted as the relative probability for producing at least one minijet pair 
at impact parameter $b$ with $p_\trans \geq p_\trans^c$ and center of mass energy $\sqrt{s}$.

For a wide range of energies, a reasonable 
parameterization of the two-gluon form factor 
is the model of Frankfurt, 
Strikman, and Weiss (FSW) \cite{Frankfurt:2003td},  
\begin{equation}
\label{eq:ggff}
{ F}_g(x,t,\mu) = \frac{1}{\left( 1 - \frac{t}{m_g^2(x,\mu)} \right)^2} \; .
\end{equation}
Spreading of the form factor at small-$x$ and evolution 
with the hard scale $\mu$ are taken into account by allowing
$m_g$ to vary with $x$ and $\mu$.  For this, we use the parameterization 
given in \cite{Frankfurt:2003td} (see Appendix B).  Performing the Fourier transform in    
Eq.~(\ref{eq:fftransform}) while using~(\ref{eq:ggff}) yields
an analytic expression for the impact parameter space gluonic form factor,
\begin{equation}
\label{eq:ggbspace}
{\cal F}_g(x,\rho,\mu) = \frac{m_g^3(x,\mu) \rho}{4 \pi} K_1 (m_g(x,\mu) \rho). 
\end{equation}  
We make one further approximation to 
simplify the analysis.   The dependence 
of $m_g$ on $x$ is rather slow.  Therefore, we make the replacement,
\begin{equation}
\label{eq:approx1}
P_2(b,x_1,x_2,p_\trans^{c }) \longrightarrow P_2(b,\bar{x},\bar{x},p_\trans^{c }) \equiv P_2(b,s,p_\trans^c) \, ,
\end{equation}
where $\bar{x} \equiv 2 p_\trans^c/\sqrt{s}$.
We have calculated $P_2(b,x_1,x_2,p_\trans^{c })$ numerically, 
and have verified that it differs from $P_2(b,s,p_\trans^c)$ by no
more than ten percent within the essential region of the integrand
in Eq.~(\ref{eq:nsemcoll2}).
We can then approximate,
\begin{equation}
\label{eq:sigsimple}
{\cal N}_2(b,s,p_\trans^c) \approx \sigma_{2 jet}^{inc}(s,p_\trans^c) P_2(b,s,p_\trans^c) \, .
\end{equation}
$P_2(b,s,p_\trans^c)$ can then be determined analytically from Eq.~(\ref{eq:ggbspace}):
\begin{equation}
\label{eq:P2}
P_2(b,s,p_\trans^c) = \frac{m_g^2(\bar{x},p_\trans^c)}{12 \pi} \left( \frac{m_g(\bar{x},p_\trans^c) b}{2} \right)^3 K_3 (m_g(\bar{x},p_\trans^c) b). 
\end{equation} 
This quantity is normalized to unity  
$\int d^2b P_2(b,s)=1$ and represents the probability density in 
impact parameter space for the minijet 
production.  Thus, the average squared impact 
parameter for minijet production is,
\begin{equation}
\label{eq:bave}
\langle \, b^2\rangle = \int d^2 {\bf b} \,\, b^2 \, P_2(b,s,p_\trans^c)\, .
\end{equation}
This provides a quantitative measure 
of the width of the overlap function which can then be used to compare
with other models.

\section{Unitarity Constraints}

We address the unitarity issue directly 
by maintaining the impact parameter representation and by calculating, 
within a particular model, the relative contribution to the total inelastic
cross section from the exclusive dijet and $2k$-jet production cross sections. 
In this section, we start with a simple model with uncorrelated jet production 
at large impact parameters.

\subsection{Multiple Collisions}

We follow steps similar to~Ref.~\cite{Ametller:1987ru} to take into account multiple collisions.
In the simple parton model picture, each collision is 
a two-to-two collision between partons resulting in an outgoing minijet pair.
For production of $k$ minijet pairs, with $k \geq 1$, there is 
a convolution integral analogous to Eq.~(\ref{eq:nsemcoll2}) which 
we symbolize by ${\cal N}_{2k}(b,s)$.  (Multiple jet production 
will also involve cutoffs on transverse momentum.)  
For $k$ minijets, the normalization condition is,
$$
\label{eq:intN}
\int d^2 {\bf b} \, {\cal N}_{2k}(b,s) \, = \, \sigma^{inc}_{2k}(s) \; ,
$$
where $\sigma^{inc}_{2k}$ is the $2k$-jet inclusive cross section.
We will refer the ${\cal N}_{2k}(b,s)$ as ``impact parameter cross sections'' since
they integrate to total inclusive cross sections for producing $2 k$ minijets.
We have not used the symbol $\Gamma(s,b)$ since this is already used 
to represent the profile function Eq.~(\ref{eq:pp_profile}) which integrates to the total cross section. 
Next, we define the ${\cal \tilde{N}}_{2k}(b,s)$ ($k \ge 1$) 
to be the exclusive analogues of the  ${\cal \tilde{N}}_{2k}(b,s)$.
That is,
$$
\int d^2 {\bf b} \, {\cal \tilde{N}}_{2k}(b,s) \, = \, \sigma^{ex}_{2k}(s) \; ,
$$
is the cross section for producing \emph{exactly} $k$ 
minijet pairs at impact parameter $b$.
{}From the definition of exclusive quantities,
the inclusive 2-jet impact parameter cross section can be 
expressed in terms of the exclusive impact parameter cross sections by writing,
\begin{equation}
{\cal N}_{2}(b,s,p_\trans^c) = \sum_{n=1}^{\infty} \, n\, {\cal \tilde{N}}_{2n}(b,s) \; .
\end{equation}
More generally, for inclusive $2k$ jet production we have
\begin{equation}
{\cal N}_{2k}(b,s) = \sum_{n \ge k}^{\infty} \,{n \choose k} \, {\cal \tilde{N}}_{2n}(b,s) \; .
\label{eq:sigma_inc}
\end{equation}
This formula says that to every inclusive $2k$ 
jet production process there is a $2n$ jet exclusive
contribution where we have ${n \choose k}$
combination of choosing $k$ pairs from $n$ pairs of jets.
Equation~(\ref{eq:sigma_inc}) can readily be inverted  
to find the exclusive ${\cal \tilde{N}}_{2k}(b,s)$ in terms
of the inclusive ${\cal N}_{2n}(b,s)$,
\begin{equation}
{\cal \tilde{N}}_{2k}(b,s) = \sum_{n\ge k}^{\infty} \,{n \choose k} (-1)^{n-k} {\cal N}_{2n}(b,s) \; .
\label{eq:sigma_exc}
\end{equation}
The contribution $\Gamma_{jets}^{inel}(b,s)$ from minijet pairs to the 
total inelastic cross section
is by definition the sum over the individual exclusive contributions,
\begin{equation}
\Gamma_{jets}^{inel}(s,b) =\sum_{k=1}^{\infty} {\cal \tilde{N}}_{2k}(b,s) \; .
\label{eq:jets_inelastic_exc}
\end{equation}
Using (\ref{eq:sigma_exc}) in (\ref{eq:jets_inelastic_exc}) produces the 
simple expression,
\begin{equation}
\Gamma_{jets}^{inel}(s,b) =\sum_{n=1}^{\infty} (-1)^{n-1} {\cal N}_{2n}(b,s)\; .
\label{eq:jets_inelastic_inc}
\end{equation}
$\Gamma_{jets}^{inel}(s,b)$ would be exactly 
equal to
$$
\Gamma^{inel}(s,b) = 2 \Gamma(s,b) - |\Gamma(s,b)|^2 \; ,
$$ 
if minijet production were present in all inelastic events.
In general, therefore, we have the unitarity constraint,
$$
\Gamma_{jets}^{inel}(s,b) \leq \Gamma^{inel}(s,b).
$$
Of course, to evaluate $\Gamma_{jets}^{inel}(s,b)$ one needs to 
know how to calculate the production cross section 
in impact parameter space for arbitrary number of jets, i.e. one needs
a model for each term in Eq.~(\ref{eq:sigma_exc}).

\begin{figure*}
\centering
  \begin{tabular}{c@{\hspace*{5mm}}c}
    \includegraphics[scale=0.5]{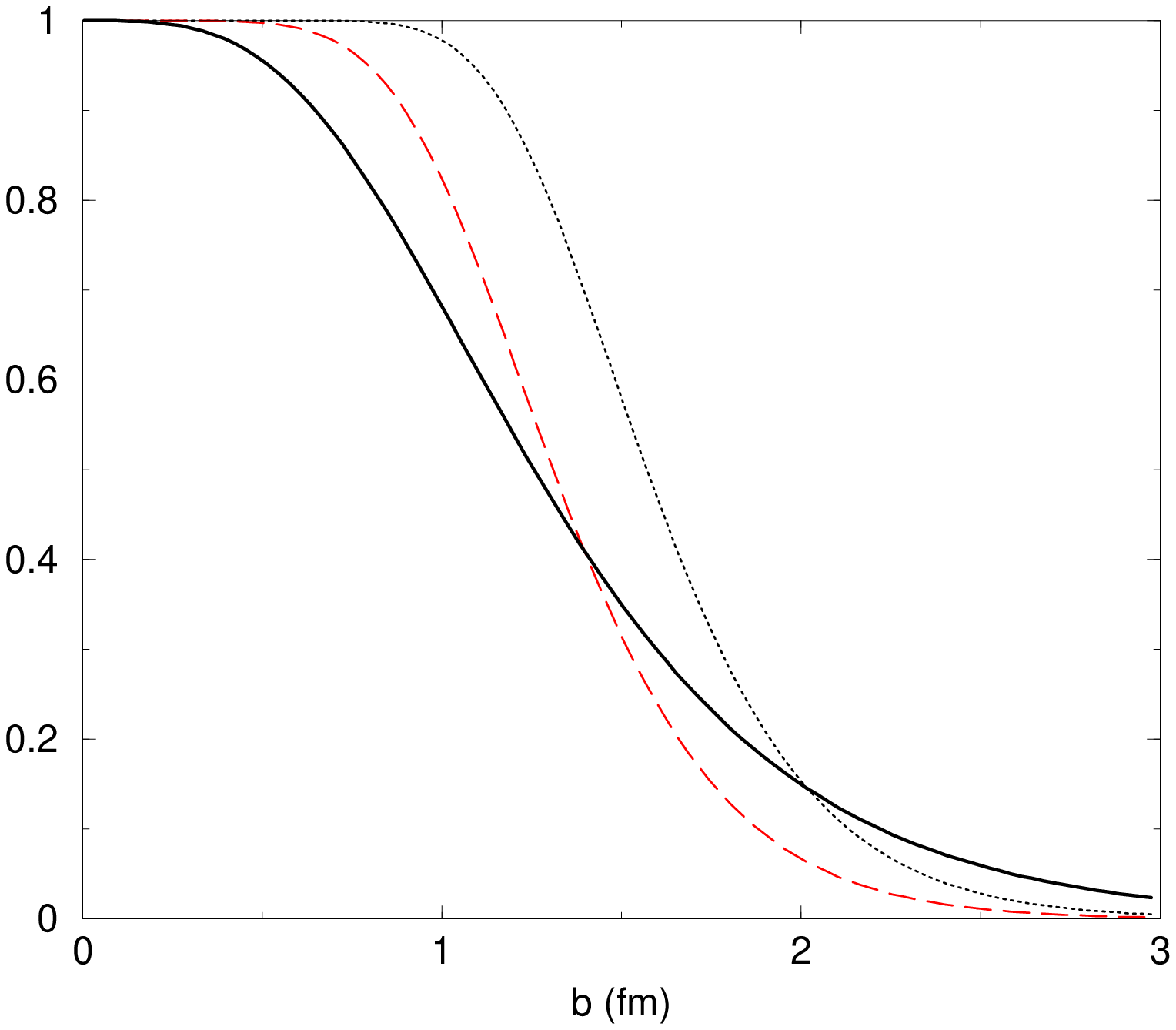}
    &
    \includegraphics[scale=0.5]{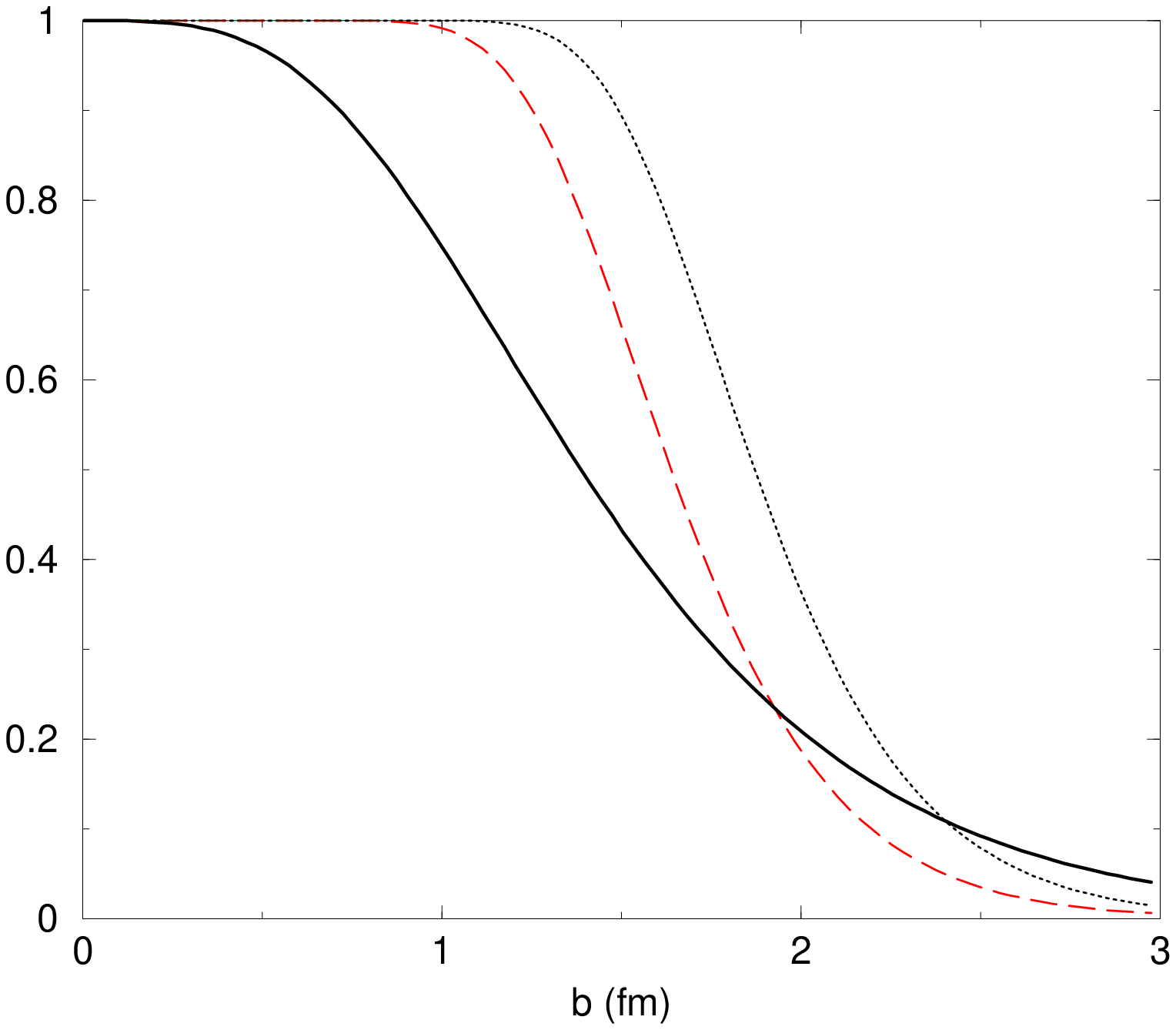}
  \\
  (a) & (b)
  \\[3mm]
    \includegraphics[scale=0.5]{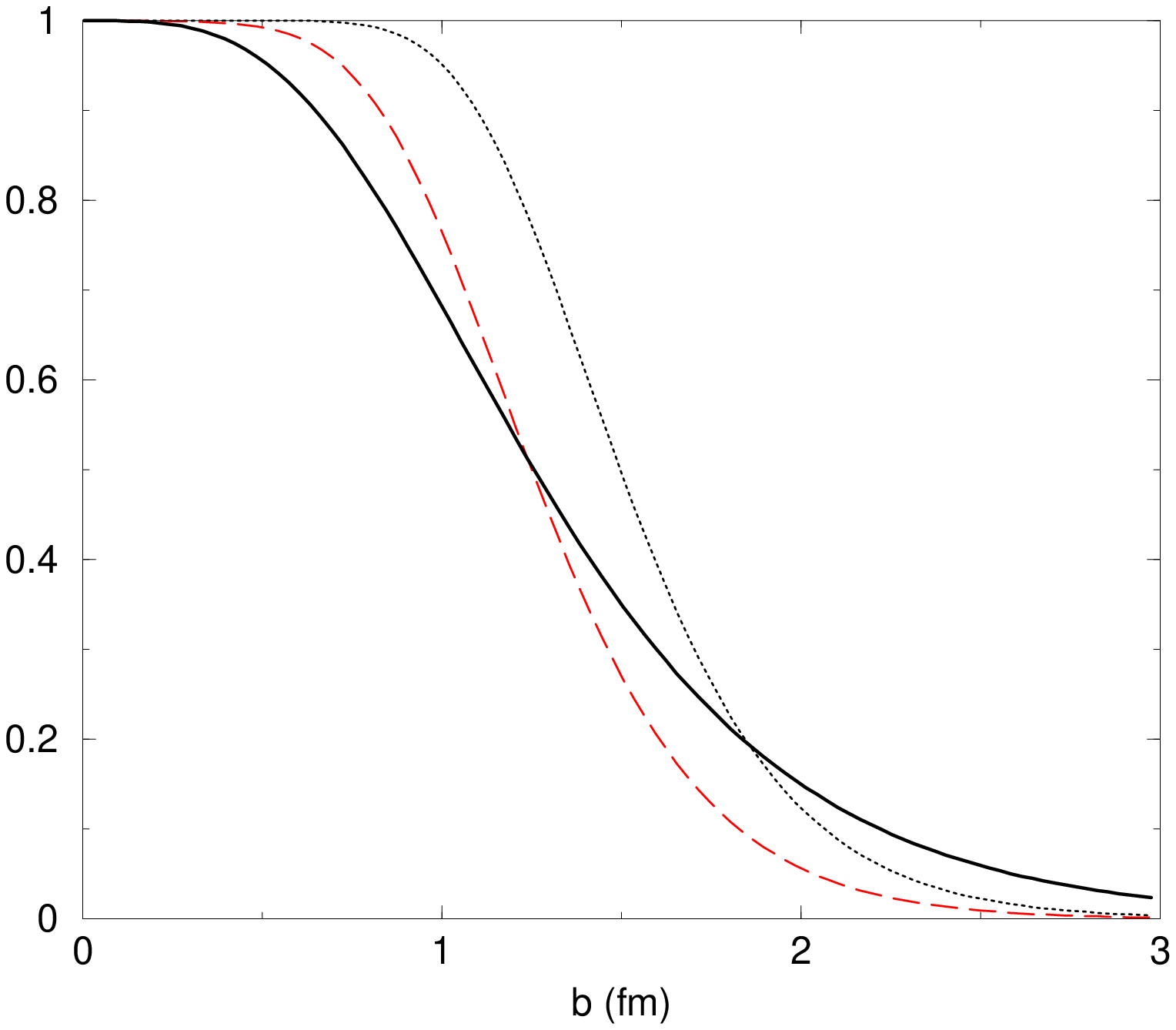}
    &
    \includegraphics[scale=0.5]{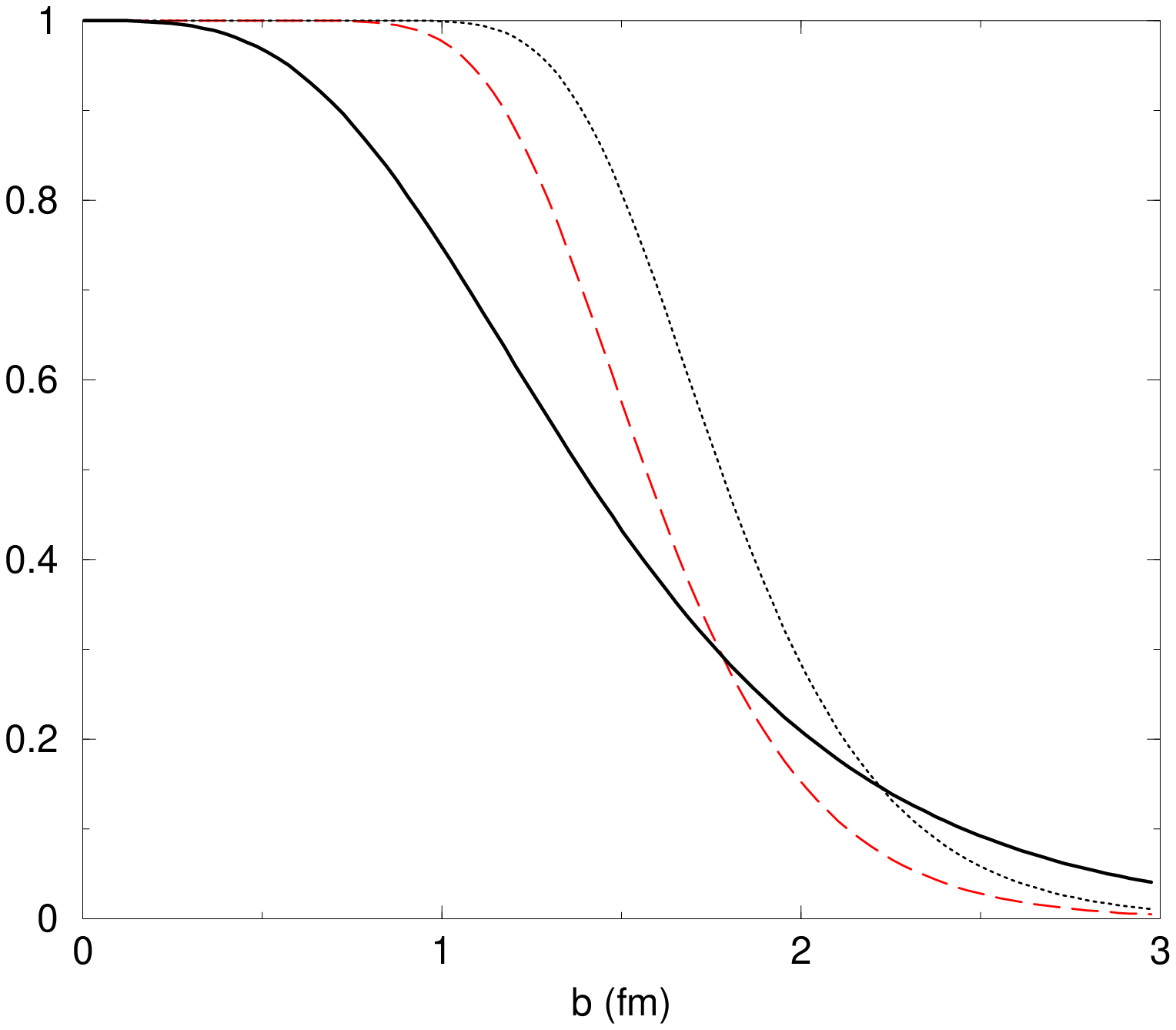}
  \\
  (c) & (d)
  \\[3mm]
  \end{tabular}
\caption{In all of these curves, 
the solid black curve is the extrapolation of the 
fit of Islam et al. (keeping only the diffractive, assymtotic limit).  
The black dotted curve is the calculation
with $p_\trans = 2.5$~GeV and the red-dashed curve is for 
$p_\trans = 3.5$~GeV.  
The upper plots are calculated using CTEQ6M parton distributions
and the lower plots are calculated used MRST parton distributions.
(a,c) $\sqrt{s} = 14$~TeV. 
(b,d)$\sqrt{s} = 50$~TeV.  
These calculations are done assuming identical partons - Eq.~(\ref{eq:geom:series_ident}).}
\label{minijet_profile}
\end{figure*}
For this paper, we adopt the simplifying approximations
of Eqs.~(\ref{eq:approx1},~\ref{eq:sigsimple}) 
of simplifying (yet reasonable) assumptions.
As already mentioned, we consider 
large impact parameters so that we may neglect correlations.
If we interpret ${\cal N}_{2}(b,\bar{x},p_\trans^c)$ as 
the probability to produce a minijet pair at impact parameter
$b$, then we may write  
the inclusive impact parameter cross section for 
producing $2k$ minijet pairs as,
\begin{equation}
{\cal N}_{2k}(s,b) ={\cal N}_{2k}(b,\bar{x},p_\trans^c) 
= (\sigma_{2jet}^{inc} P_2(b,\bar{x},p_\trans^c))^k\; .
\label{eq:themodel}
\end{equation}
The expression (\ref{eq:jets_inelastic_inc})  for a given 
value of impact parameter is then a geometric series that yields the 
analytic expression,
\begin{equation}
\Gamma_{jets}^{inel}(s,b) =\sum_{n=1}^{\infty}(-1)^{n-1} {\cal N}_{2}^n 
= \frac{\sigma_{2jet}^{inc} P_2(b,\bar{x},p_\trans^c)}{1+\sigma_{2jet}^{inc} P_2(b,\bar{x},p_\trans^c)} \; .
\label{eq:geom:series}
\end{equation}
If the final state partons are identical, then each term 
in Eq.~(\ref{eq:themodel}) should carry a symmetry factor of $1/k!$.
In that case the series in Eq.~(\ref{eq:geom:series}) becomes,
\begin{equation}
\Gamma_{jets}^{inel}(s,b)
= 1-\exp \left[ -\sigma_{2jet}^{inc} P_2(b,\bar{x},p_\trans^c) \right]\; .
\label{eq:geom:series_ident}
\end{equation}
We will investigate both of these cases.
Note that, as discussed in ~\cite{Ametller:1987ru}, Eqs.(~\ref{eq:geom:series}) and (~\ref{eq:geom:series_ident}) satisfy the unitarity 
condition $${\Gamma}_{jets}^{inel}(s,b) \le 1 \; ,$$ by construction.
Equations~(\ref{eq:geom:series}) and (\ref{eq:geom:series_ident}) are a reasonable 
approximation at values of $b$ where $N_4(b,s)/N_2(b,s)$ is a small parameter so that
corrections due to correlations are small.  It turns out that 
this is true for $b \approx 2$~fm for energies up to about $\sqrt{s} \approx 50$~TeV.

In the specific case of four jets the profile function is defined as
$$
P_4(b) \, = \, \frac{P_2(b)^2}{\int d^2 b P_2(b)^2} \, ,
$$
where we have suppressed the other arguments in the profile functions.
Note that $P_4$ satisfies normalization condition by construction.
Likewise the cross section for the four jet production is given by
$$
\sigma_{4jet} =  \sigma^2_{2jet} \, \int d^2 b P_2(b)^2 \, .
$$

\subsection{Numerical Calculations}

Figure~\ref{minijet_profile} shows sample calculations using 
Eq.~(\ref{eq:geom:series}) for the case of (a): $\sqrt{s} = 14$~TeV, relevant 
to the LHC, and for  (b): $\sqrt{s} = 50$~TeV, relevant to cosmic ray air showers. 
Here we assume that the partons are identical particles.
To test the sensitivity to the type of parton distribution being used,
we have repreated the calculation using MRST parton distributions in plots (c,d).
We have done the calculation using two values of $p_\trans$ in each case: the lower red-dashed 
curve is with $p_\trans = 3.5$~GeV, and the upper dotted curve is with $p_\trans = 2.5$~GeV.
In both of these calculations, we have used CTEQ6M gluon distribution functions.
For comparison, we have shown the profile for the total inelastic cross section obtained from 
the fits
 of Islam et al~\cite{Islam:2002au} ({It is difficult  at present to estimate the uncertainties in the 
extrapolation  of models for the inelastic impact parameter profile to the LHC energy. 
However the first  data from LHC on elastic $pp$ scattering will essentially eliminate this uncertainty.}).  
In Fig.~\ref{minijet_profile}a, the calculation with a cutoff at $p_\trans = 2.5$~GeV
results in a value for $\Gamma^{inel}_{jets}$ that is larger than the inelastic
profile function, $\Gamma^{inel}$ at $b \gtrsim 1.5$~fm, whereas the cutoff of $p_\trans^c = 3.5$~GeV
leads to a violation only at small values of $b$ where the uncorrelated approximation likely breaks
down.  Therefore, the $p_\trans = 2.5$~GeV cutoff is certainly inappropriate for the maximum LHC energies.
In Fig.~\ref{minijet_profile}b we show the calculation for the higher energy $\sqrt{s}=50 \; {\rm TeV}$. 
Here we see that the $p_\trans$ needs to be pushed even higher than $3.5$~GeV to avoid
a contradiction with unitarity. 
To test the sensitivity to the type of parton distribution being used,
we have repreated the calculation using MRST parton distributions in plots (c,d).  We see that the
above conclusions are essentially unchanged.

It is worth emphasizing  that typical values of $b$ for jet production are much smaller, $\sim 0.8 \, {\rm fm}$  
than the ones for which we obtained the constraint. 
Since  the typical gluon densities involved for $b \sim 0.8 \, {\rm fm}$ 
are significantly larger (of the order of factor $3$,  compare Eq.~(\ref{eq:ggbspace}))
 than for $b \sim 1.5 \; {\rm fm}$, the taming mechanism should be effective in the much broader range of $p_t$.

This result supports the cutoffs used in DPMjet II and III~\cite{DPMjet} in which the cutoff
at this energy scale is roughly $6$~GeV, and SIBYLL 2.1~\cite{SIBYLL} in which the cutoff is 
roughly $5$~GeV
independent of $b$.  However, in our logic it seems natural for the cutoff to inrease with a decrease of $b$.

Our results disagree with the use of a low fixed 
cutoff as in QGSjet~\cite{QGSjet} where the minimum $p_\trans$ is fixed at $1.5$~GeV,
and neXus~\cite{nexus} where the cutoff is fixed at $2.0$~GeV.  
Note that it is to natural to expect that the $p_\trans^2$ cutoff 
is proportional to average gluon density encountered for  
a given impact parameter. Hence our limit for $b \sim 1.5$~fm 
corresponds to a much larger value of the cutoff for 
more typical $b$-values for the hard collisions - 
say production of two jets where average $b$ are $\sim 1$~fm. 

\begin{figure*}
\centering
  \begin{tabular}{c@{\hspace*{5mm}}c}
    \includegraphics[scale=0.5]{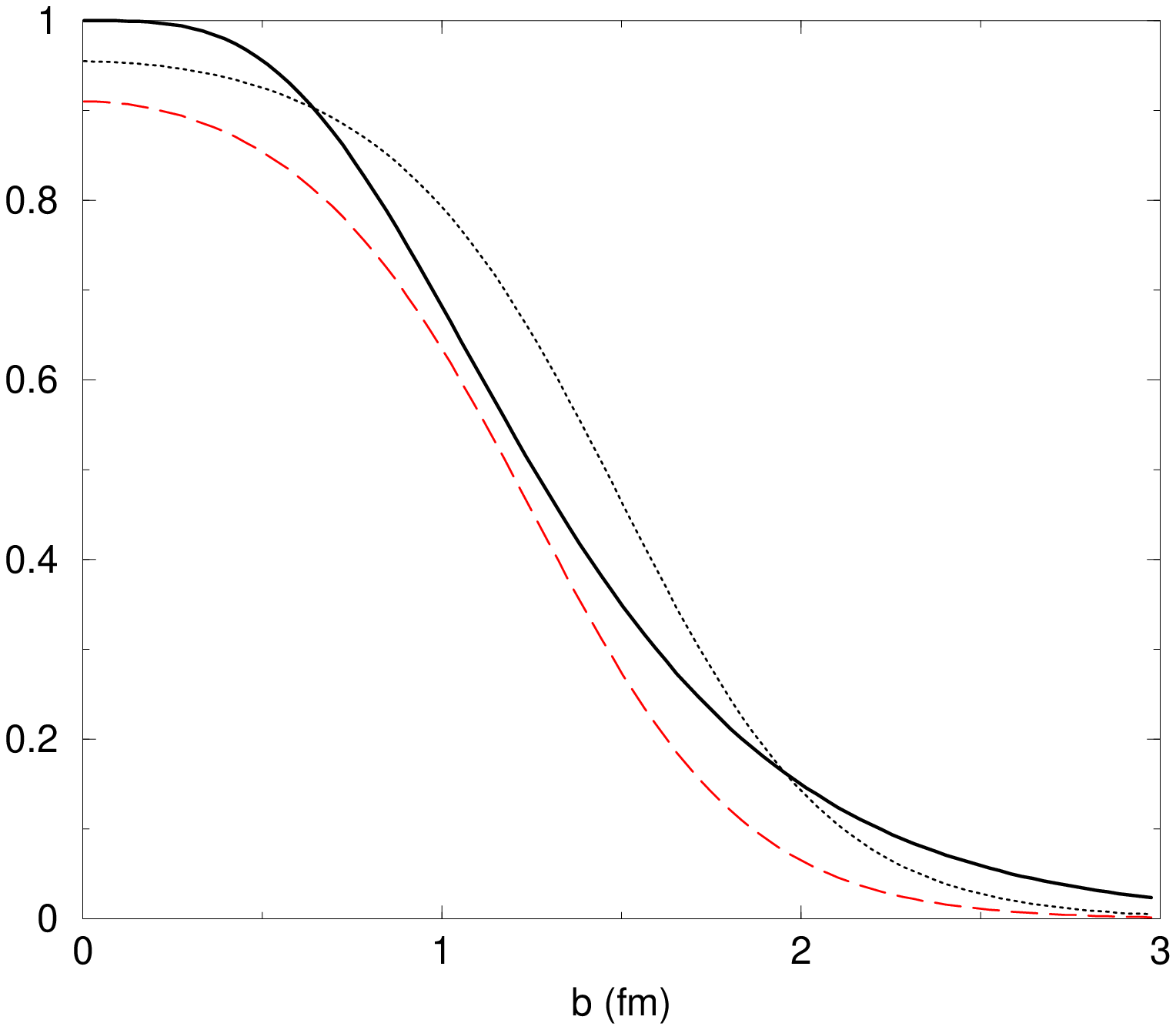}
    &
    \includegraphics[scale=0.5]{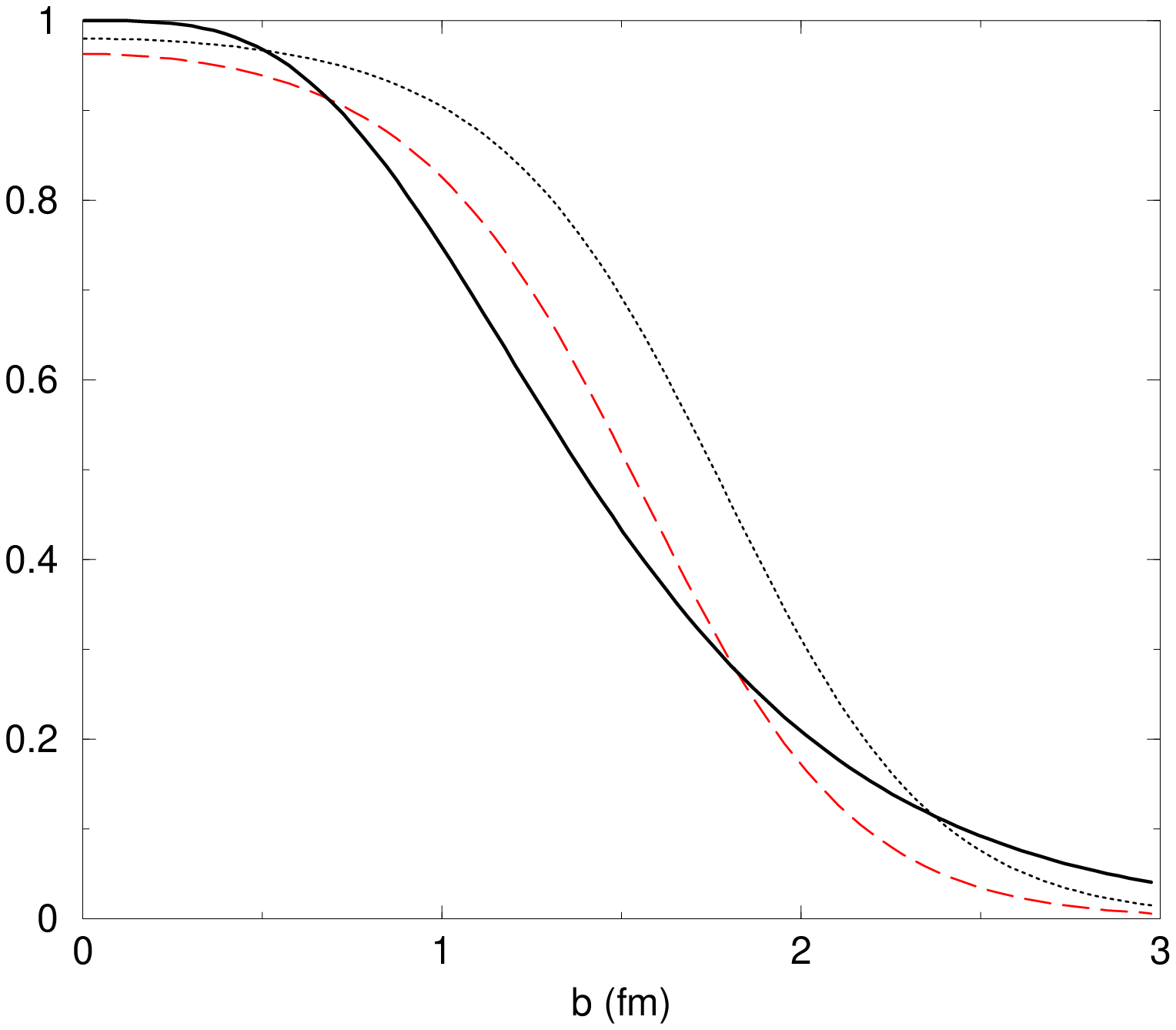}
  \\
  (a) & (b)
  \\[3mm]
    \includegraphics[scale=0.5]{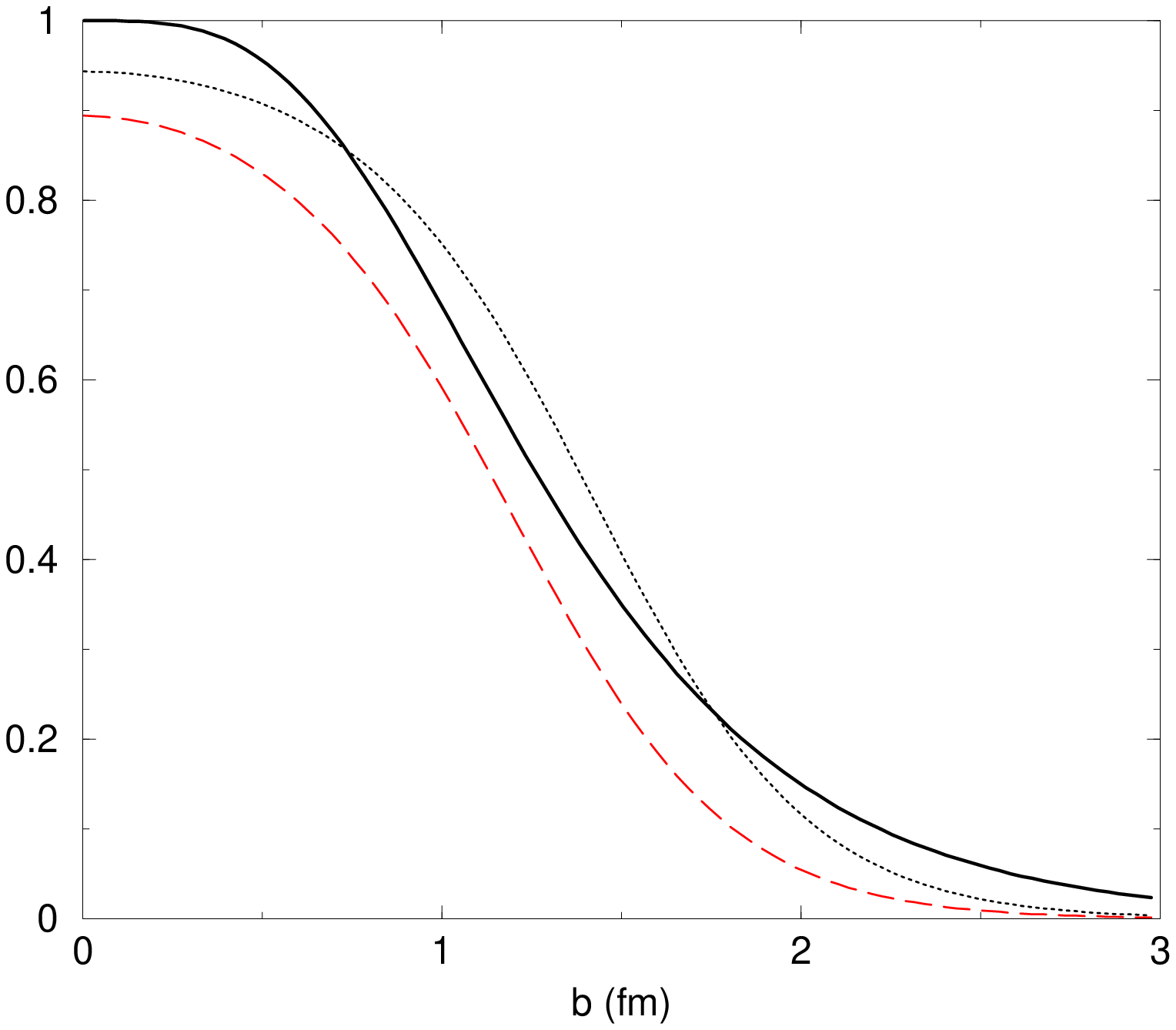}
    &
    \includegraphics[scale=0.5]{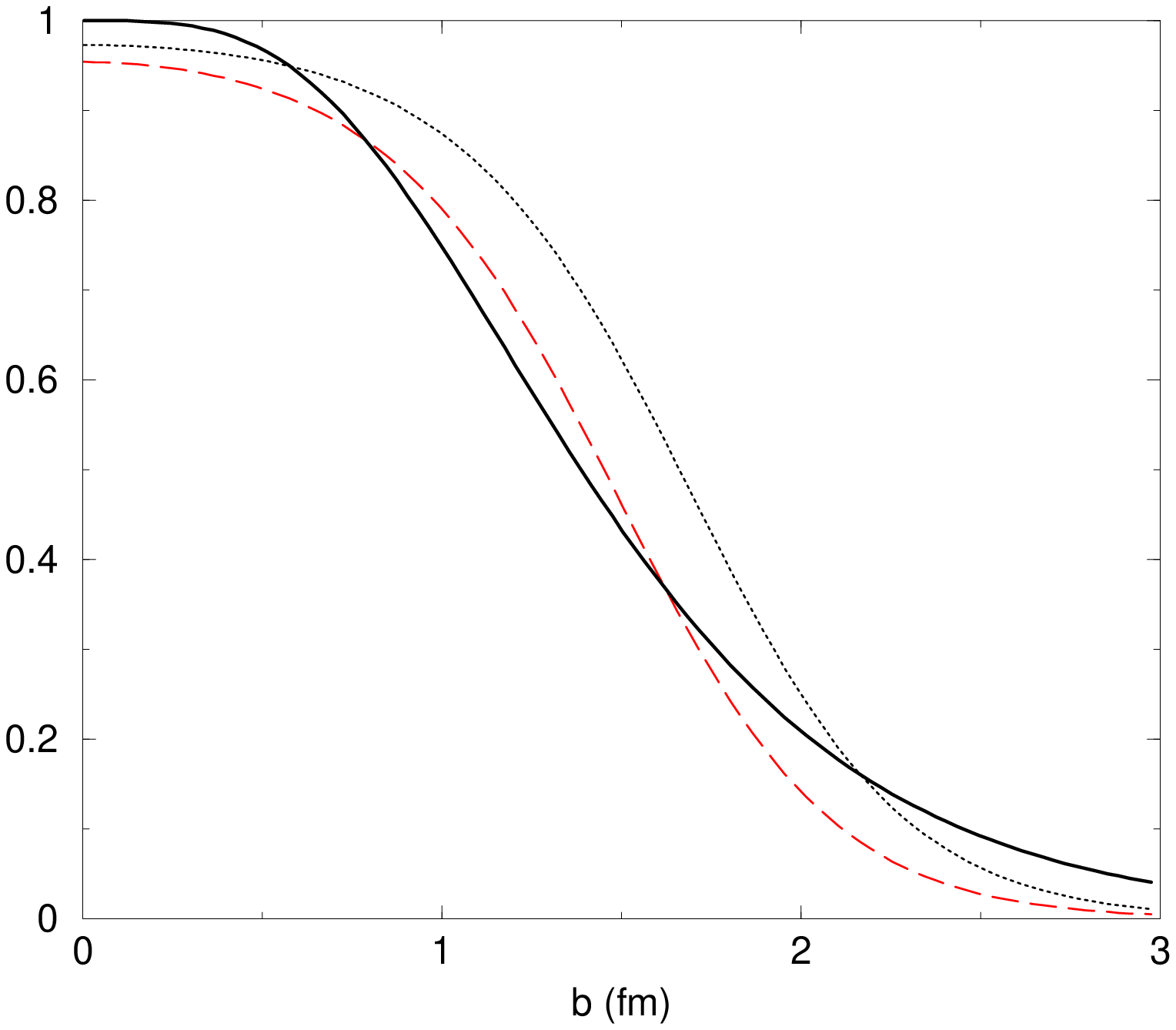}
  \\
  (c) & (d)
  \\[3mm]
  \end{tabular}
\caption{In all of these curves, 
the solid black curve is the extrapolation of the 
fit of Islam et al. (keeping only the diffractive, asymptotic limit).  
The black dotted curve is the calculation
with $p_\trans = 2.5$~GeV and the red-dashed curve is for 
$p_\trans = 3.5$~GeV. 
The upper plots are calculated using CTEQ6M parton distributions
and the lower plots are calculated used MRST parton distributions.
Furthermore, the calculation 
is done assuming non-identical partons - Eq.~(\ref{eq:geom:series}).
(a,c) $\sqrt{s} = 14$~TeV. 
(b,d)$\sqrt{s} = 50$~TeV.}
\label{minijet_nonidentical}
\end{figure*}

We have also repeated the calculation using the 
assumption of non-identical partons - Eq.~(\ref{eq:geom:series}).  
This is shown in Fig.~\ref{minijet_nonidentical}.  We see 
that our above conclusions regarding the minimum $p_\trans$  
are not substantially affected.

\begin{figure*}
\centering
  \begin{tabular}{c@{\hspace*{5mm}}c}
    \includegraphics[scale=0.5]{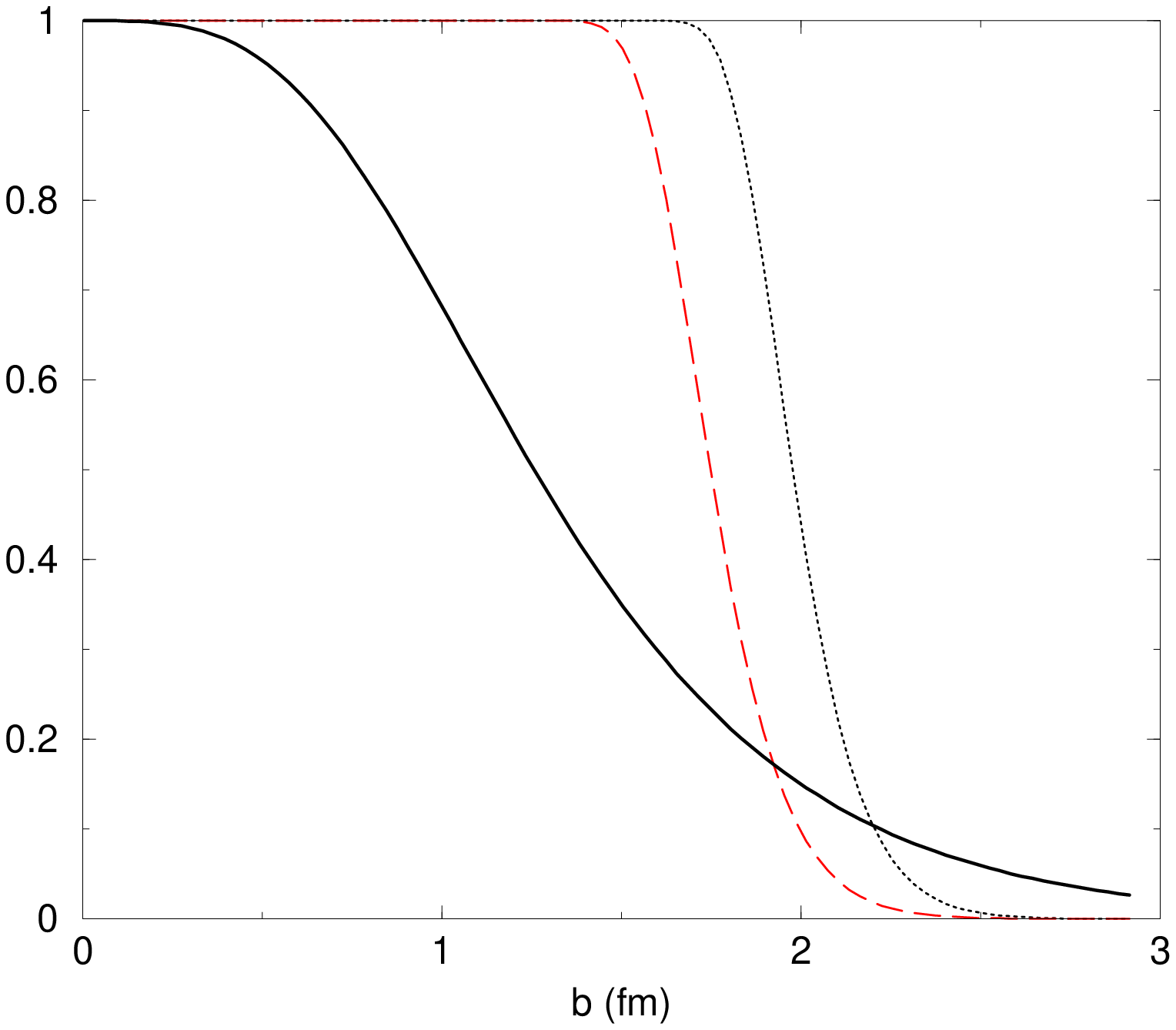}
    &
    \includegraphics[scale=0.5]{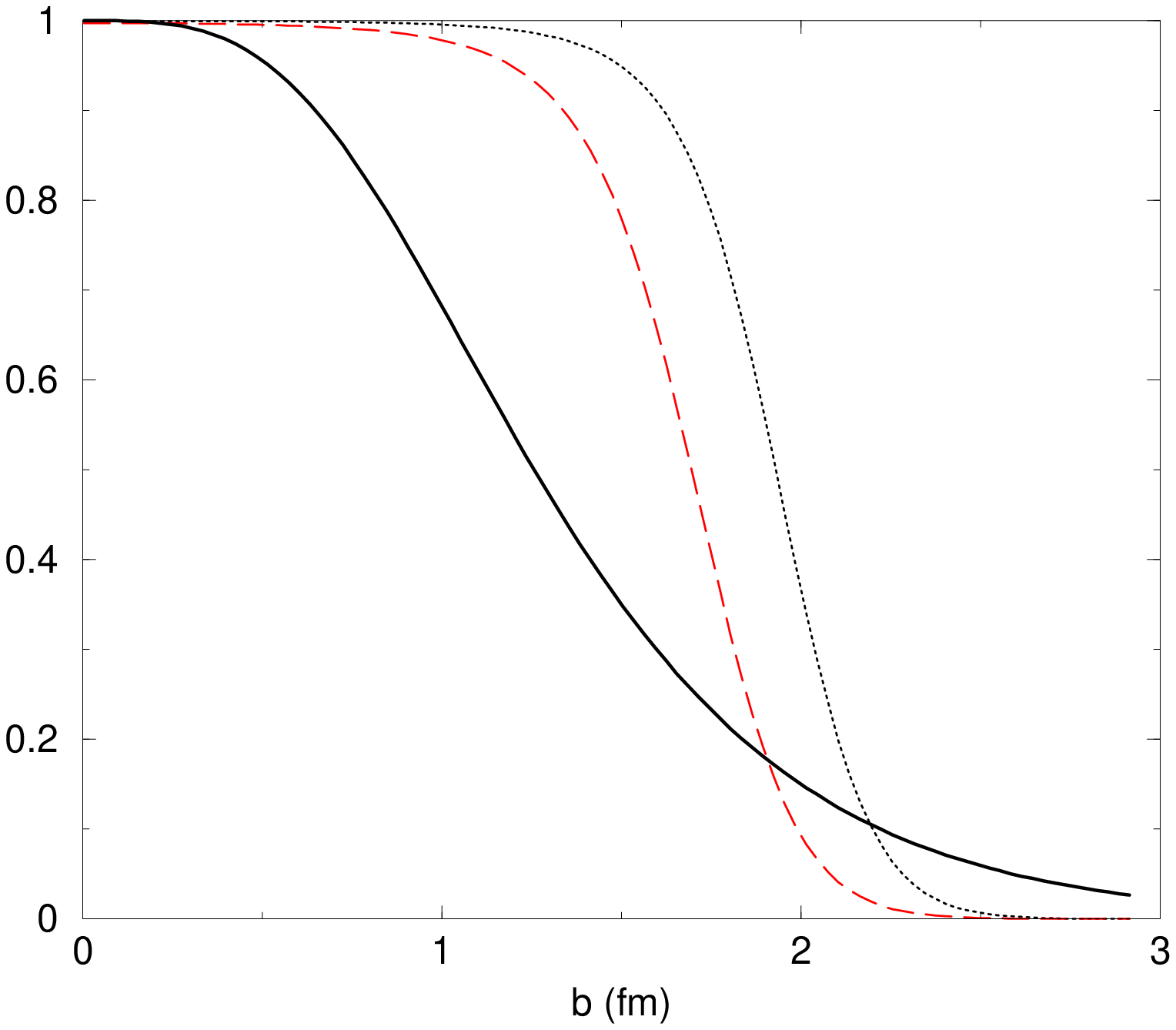}
  \\
  (a) & (b)
  \\[3mm]
  \end{tabular}
\caption{For these curves we use a 
Gaussian model for $P_2$.  Here $\sqrt{s} = 14$~TeV and 
we use CTEQ6M parton distributions.
(a) Calculation for identical partons.  (b) For non-identical partons.}
\label{minijet_gaussian}
\end{figure*}

One source of uncertainty is in the true shape of the $t$-distribution of the 
two-gluon form factor.
To test the sensitivity to this, we have repeated the analysis of the previous 
section using the more commonly used Gaussian model for the $t$-dependence of $P_2$ rather than 
the two-gluon form factor.  We fix the slope by demanding that $\langle b^2 \rangle$ for 
the Gaussian parameter equal the value of $\langle b^2 \rangle$ determined from the 
two-gluon form factor.
The result is shown in Fig.~\ref{minijet_gaussian}. 
In this case, it seems that using the Gaussian parameterization leads to a greater
violation of unitarity than the GPD.


\section{Interaction Radius at Very High Energies}
\label{sec:eikonal_label}

In this section we give a more general discussion of 
semi-hard interaction radius.  We argue that it should
be at least as large as what is used in the previous section 
to maintain consistency with DIS.
A common approach to modeling the impact parameter dependence 
in extrapolations to very high energies is 
to write the profile function as,
\begin{equation}
\Gamma(s,b) = 1 - \exp \left[-\chi(s,b) \right]. \label{eq:gamma}
\end{equation}
Here $\chi(s,b)$ is assumed to describe the phase 
shift produced by each scattering of the proton constituents.

In extrapolations to ultra-high energies it is also assumed that 
the basic parton-parton profile function can be decomposed into the sum of a 
term for soft scattering and for hard scattering:
\begin{equation}
\label{totprofile}
\Gamma(s,b) =  1 - \exp \left\{ - \chi_{h}(s,b) - \chi_{s}(s,b) \right\}\; .
\end{equation}
Corrections to this simple model are taken into account 
in more sophisticated versions of the eikonal model
that include, for example, diffraction and triple 
Pomeron exchange~\cite{DPM}, though for the remaining discussions 
of this section, it will be sufficient to use Eq.~(\ref{totprofile}).
The soft part $\chi_s(s,b)$ is modeled phenomenologically, 
whereas the semi-hard term $\chi_h(s,b)$ can 
be calculated with the aid of Eq.~(\ref{minijet}).
However, we stress that since each term in the exponent of Eq.~(\ref{eq:gamma})
is quite model dependent the relationship between the unitarity 
of the total cross section and the large size of the semi-hard 
contribution, $\chi_h(s,b)$ is unclear, and is likely to depend
strongly on how we model the soft parts, etc.  
One can extract constraints on the UHE cross section indirectly by noting that
the inclusive cross sections are 
obtained via sums over exclusive 
$2n$ minijet cross sections,
\begin{equation}
\label{sumexclusive}
\sigma_{tot} = \sum_{n=0}^{\infty} \tilde{\sigma}_{2n}, \qquad 
\sigma_{2jet}^{inc} = \sum_{n=1}^{\infty} n \, \tilde{\sigma}_{2n}\; .
\end{equation}
(Here it is assumed that the total cross section is dominated by jet pairs.)
As in the previous section, we use a tilde on exclusive quantities.
The total and elastic cross sections are, from Eqs.~(\ref{eq:sigtot}) and~(\ref{eq:sigel}),
\begin{eqnarray}
\label{eq:tot_ela}
\sigma_{\rm tot} = 2 \int d^{2} {\bf b} \, \left[ 1 - \exp \left\{ - \chi_{h}(b) - \chi_{s}(b) \right\} \right] \label{totcros} \; , \\
\sigma_{\rm ela} = \int d^{2} {\bf b} \, \left[ 1 - \exp \left\{ - \chi_{h}(b) - \chi_{s}(b) \right\} \right]^{2} \; .
\end{eqnarray}
In~\cite{Engel:2001mm}, for example, a minimum value for $p_\trans^c$ was determined
by requiring fits of $\sigma_{\rm tot}$ and 
$\sigma_{\rm ela}$ to data for the elastic cross section to be consistent with
Eqs.~(\ref{sumexclusive}).  In the semi-hard eikonal factor, a
simple Gaussian model is used,
\begin{equation}
\label{hardprofile}
\chi_{h}(b) = \frac{\sigma_{h}(s)}{8 \pi R_0^2} \exp \left\{ -b^{2}/4 R_0^2 \right\}. 
\end{equation}
$R_{0}$ is the interaction radius in impact parameter space at 
the reference energy scale.  (In general, there should 
also be a non-zero Regge slope $\alpha^{\prime}$ leading
to diffusion  in the transverse plane as energy is increased.)  
\subsection{Comparison with Deeply Inelastic Scattering}
\label{dipole_unitarity}

{}From Eq.~(\ref{eq:N_norm}) and Eq.~(\ref{hardprofile}) we can extract
the equivalent of Eq.~(\ref{eq:P2}) for the Gaussian model,
\begin{equation}
\label{eq:gauss_overlap}
P_2^{gauss}(b,s,p_\trans^c) = \frac{1}{4 \pi R_0^2} \exp \left\{ -b^{2}/4 R_0^2 \right\}.
\end{equation}
When fitted to data the two over-lap functions Eq.~(\ref{eq:gauss_overlap}) and Eq.~(\ref{eq:P2}) 
should be in rough agreement. 
In particular, both models should yield
similar numerical values for $\langle \,b^2\rangle$ calculated with Eq.~(\ref{eq:bave}).
For example, if Eq.~(\ref{hardprofile}) is used with a radius, $R_0 = 3.5$ GeV$^{-2}$ 
and a transverse momentum cutoff of $p_\trans^c = 3.5$~GeV, one obtains a mean impact parameter
$\sqrt{\langle \, b^2\rangle}$ of $0.75$~fm, whereas 
if Eq.~(\ref{eq:P2}) is used one obtains $\sqrt{\langle \, b^2\rangle} = 0.87$~fm.
In this case there is at least rough agreement between the two models.

\begin{figure}
\epsfig{file=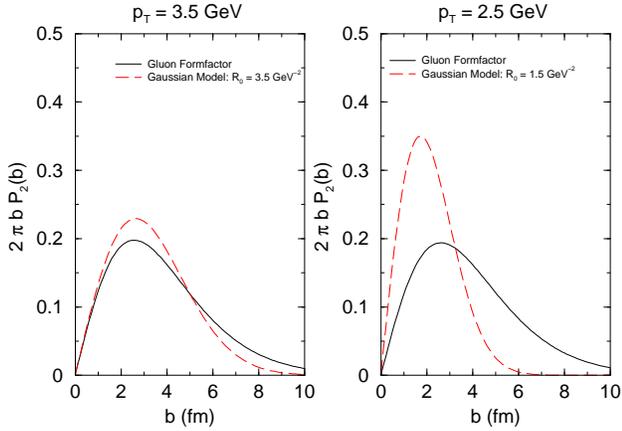,scale=.5}
\caption{A comparison of the distribution in impact parameter space
for minijet production as predicted by the Gaussian model, Eq.~(\ref{eq:gauss_overlap}) 
and the 2-gluon form factor, Eq.~(\ref{eq:P2}). }
\label{profilesm}
\end{figure}

In Ref.~\cite{Engel:2001mm} it is found that
a fit to the total cross section is also possible 
with a radius of $R_0 = 1.5$ GeV$^{-2}$ and $p_\trans^c = 2.5$~GeV.
However, in this case Eq.~(\ref{hardprofile}) produces a mean impact parameter equal to
$\sqrt{\langle \, b^2\rangle}$ of $0.48$~fm, whereas using 
Eq.~(\ref{eq:P2}) produces $\sqrt{\langle\,  b^2\rangle} = 0.89$~fm.
For this scenario, there is clearly an inconsistency between the 
Gaussian model and the impact parameter dependence extracted from DIS.
The spread in impact parameter space for these two scenarios is shown in Fig.~\ref{profilesm}.

We can gain further insight 
by examining how the width of the 
impact parameter dependence of ${\cal F}_g(x,\rho,\mu)$ 
relates to the total $q \bar{q}$-proton cross section.
In DIS at high energy (or low-$x$), 
cross sections are usually written in the target rest frame in term of the cross 
section for the interaction of a $q\bar{q}$ dipole with 
the target proton.  The $q \bar{q}$ cross section is,
\begin{equation}
\label{dipole}
\sigma^{q \bar{q}}_{tot}(d,x) = \frac{\pi^{2}}{3} d^{2} \alpha_{s} (\lambda/d^{2}) x g(x,\lambda/d^{2})\,,
\end{equation}
where $d$ is the transverse size of the $q \bar{q}$ pair, and $x$ is 
the longitudinal momentum fraction carried by the struck gluon\footnote{This formula is derived for small $d$ when $\sigma_{tot} \approx
\sigma_{inel}$}.
Given the total cross section for the $q \bar{q}$-proton interaction
and the GPD in Eq.~(\ref{eq:ggff}), we can write the amplitude for dipole scattering as,
\begin{equation}
\label{basicamp}
A_{q\bar{q}}(x,t) = i \hat{s} \sigma^{q\bar{q}}_{tot}(d,x) { F}_g(x,t,\mu)\,, 
\end{equation}
Since we are in the high energy limit, 
we neglect a small real part.  $\sqrt{\hat{s}}$  is the center-of-mass 
energy of the dipole-proton collins.  
We then invert Eq.~\ref{basicamp} to obtain a profile function 
for the $q \bar{q}$-proton cross section for a pair of size $d$ and for a gluon momentum fraction, $x$:
\begin{multline}
\label{hardprofile2}
\Gamma^{q\bar{q}}_{h}(b,x) = \\ \frac{\sigma^{q \bar{q}}_{tot}(d,x) 
m_g^2(x,\mu)}{4 \pi} \left( \frac{m_g(x,\mu) b}{2} \right) K_1 (m_g(x,\mu) b)\,.
\end{multline}

In Eq.~(\ref{minijets}), the hard scale at which the 
gluon density and the strong coupling are evaluated is $p_\trans$ 
whereas in the DIS expression, Eq.~(\ref{dipole}),
it is related to the inverse size of the dipole's transverse size, $\lambda/d^{2}$,
where $\lambda$ is between $4$ and $10$\cite{Rogers:2003vi}.  Because of the universality 
of the GPD, the gluon 
density in Eq.~(\ref{dipole}) should be the same as that appearing in Eq.~(\ref{minijets}) so long
as we consider a dipole size $d$ such that $\lambda/d^{2} \sim p_\trans^{2}$.
We are interested in the region of the integral close to $p_\trans^c$ and 
for typical values of $x \approx \bar{x}$.  

\begin{figure}
\epsfig{file=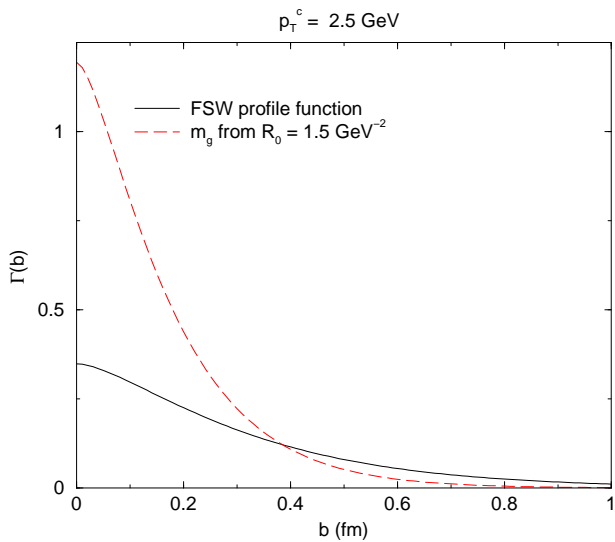,scale=.5}
\caption{The profile function (solid curve) calculated using
Eq.~\ref{hardprofile2}.  To match with the Gaussian model, $m_g$ must 
be increased by a factor of 1.9 - resulting in the dashed curve.}
\label{jpsiprof}
\end{figure}

Therefore, we evaluate Eq.~(\ref{hardprofile2}) 
at $d_c = \sqrt{\lambda/(p_\trans^{c})^2}$.
Using the FSW expression provided in Appendix B for 
$m_g(\bar{x},p_\trans^c)$ yields the solid curve shown in Fig.~\ref{jpsiprof}.
The profile function calculated this way is well within the bounds 
of the unitarity constraint which is consistent with observations at HERA.

However, for $\langle \,b^2\rangle$ to agree with the prediction from the 
Gaussian model with $R_0 = 1.5$~GeV$^{-2}$ and $p_\trans^c = 2.5$~GeV,
we find that we must increase the value of $m_g$ by about a factor of 1.9.
This leads to the dashed profile function shown in Fig.~\ref{jpsiprof} which 
exceeds the unitarity bound at small values of $b$.
{}From Fig.~\ref{profilesm} 
we see that, for the 
DIS calculation to remain consistent with unitarity,
the $p_\trans$ cutoff in the hadron-hadron jet 
cross section Eq.~(\ref{minijets}) should actually be pushed \emph{higher}
as the profile for hard partons becomes narrower.  In other words, a narrow 
profile for hard scattering leads to a more rapid approach to saturation-like 
physics in the DIS calculation.  
To avoid a conflict with the unitarity in DIS, and to 
avoid a contradiction with the observation at HERA that the 
amplitude for the dipole scattering is relatively far from the 
black disk limit  even when  $x=10^{-4} \sim 10^{-3}$ (at least 
unless the scales are greater than $\sim 2 \; {\rm GeV}^2$, 
see \cite{Munier:2001nr}) we must use a radius-squared for the 
hard interaction that is greater than $\sim 3.5$~GeV$^{-2}$. 

\section{Conclusions}

In this paper we have analyzed the region of applicability of 
the pQCD formula for the minijet production in hadron-hadron 
collisions. Based on the unitarity arguments in impact parameter 
space we have set a constraint on the minimum value of $p_\trans$ for 
which the formula can be used. The model satisfies the unitarity 
by construction, i.e. by the summation of the multiple scatterings, 
without any correlations in impact parameter space. 
For the LHC-scale 
energy 
$\sqrt{s}=14\;{\rm  TeV}$
the minimum value of $p_\trans^c \gtrsim  2.5 {\rm GeV}$ and for the cosmic 
ray energies for $pp$ collisions $\sqrt{s}=50\;{\rm  TeV}$, $p_\trans \gtrsim 3.5 \;{\rm GeV}$.

We again stress that these values  are for large impact parameters $\sim 1.5 \; {\rm fm}$.
Though our simple analysis is not effective at small impact parameters
due to the neglect of correlations and due to $\Gamma(s,b)\sim 1$, it is reasonable to expect the 
minimum $p_\trans^c$ to rise as the gluon density rises.  Thus, the true 
minimum $p_\trans^c$ is probably even larger than what we find here.

For the case of the cosmic ray interaction of protons with air, the typical gluon densities are at least
factor five larger than the gluon densities encountered in the $pp$ case at $b\sim 1.5 \;{\rm fm}$.

Note that the values of the minimum $p_\trans$ are much larger than  any conceivable
soft scale. Therefore, the dynamical mechanism for violation of the leading twist formalism
which we find here
must include some kind of non-linear strong field effects at relatively small coupling.

The analysis can be 
extended for a more sophisticated model which includes the 
correlations for multi-jet production in the impact parameter space.

We have also analyzed in detail the dominant regions of the integrand 
in the minijet formula and shown that the dominant configurations for 
small energies are when the jets are produced with approximately equal but opposite rapidities.
 When the 
energy is increased the configurations with the jets going 
into the same rapidity hemisphere are equally important.
This includes the forward regime in hadron scattering where the saturation corrections in the gluon
density are likely to be important at the LHC energy.

\section*{Acknowledgments}
We thank Ralf Engel, Leonid Frankfurt and Daniele Treleani   for discussions. We also thank Werner Vogelsang for calculations of the $K$ factor in jet production.  
This research has been supported by the U.S. D.O.E. under grants number DE-FG02-90ER-40577,
DE-FGO2-93ER-40771 
and by the Polish Committee for Scientific Research grant No. KBN 1 P03B 028 28.

\section*{Appendix A}
\subsection*{ Parametrization of the mass parameter $m_g$ in the profile function}
The mass parameter $m_g$ in (\ref{eq:ggff}) is related  to the inverse of the average impact parameter
corresponding to the profile distribution (\ref{eq:ggbspace})
$$
\langle\rho^2\rangle = \frac{8}{m_g^2}\; .
$$
In \cite{Frankfurt:2003td} the following parametrization was found for the $x$ and scale dependence of $m_g$
$$
\langle \rho^2 \rangle (x,Q_0^2) = {\rm max}\bigg\{0.31 {\rm fm}^2+0.0194 {\rm fm}^2 \ln \frac{0.1}{x},0.28 {\rm fm }^2 \bigg\}\, ,
$$
and
$$
\langle \rho^2 \rangle(x,Q^2)=\langle \rho^2 \rangle (x,Q_0^2)\bigg(1+A \ln \frac{Q^2}{Q_0^2}\bigg)^{-a} \, ,
$$
where
$$
Q_0^2 = 3 \,{\rm GeV}^2, \; \;A=1.5, \;\;a=0.0090 \ln \frac{1}{x}\;.
$$

\end{document}